# Reversible Structural Transition of Two-Dimensional Copper Selenide on Cu(111)

Yuan Zhuang,[1#] Yande Que,[1,#*] Chaoqiang Xu,[1,2] Bin Liu,[1] and Xudong Xiao[1,3*]

[1]*Department of Physics, The Chinese University of Hong Kong, Shatin, Hong Kong, China*

[2]*Postdoctocral Research Station, Shenzhen Capital Group Co. Ltd., Shen Zhen 518048, China*

[3]*School of Physics and Technology, Wuhan University, Wuhan 430072, China.*

[#] *These authors contributed equally.*

[*] Corresponding author.

Email: ydque@phy.cuhk.edu.hk , and xdxiao@phy.cuhk.edu.hk

**ABSTRACT:**

Structural engineering opens a door to manipulating the structures and thus tuning the properties of two-dimensional materials. Here, we report a reversible structural transition in honeycomb CuSe monolayer on Cu(111) through scanning tunneling microscopy (STM) and Auger electron spectroscopy (AES). Direct selenization of Cu(111) gives rise to the formation of honeycomb CuSe monolayers with 1D moiré structures (stripe-CuSe), due to the asymmetric lattice distortions in CuSe induced by the lattice mismatch. Additional deposition of Se combined with post annealing results in the formation of honeycomb CuSe with quasi-ordered arrays of triangular holes (hole-CuSe), namely, the structural transition from stripe-CuSe to hole-CuSe. Further, annealing the hole-CuSe at higher temperature leads to the reverse structural transition, namely from hole-CuSe to stripe-CuSe. AES measurement unravels the Se content change in the reversible structural transition. Therefore, both the Se coverage and annealing temperature play significant roles in the reversible structural transition in CuSe on Cu(111). Our work provides insights in understanding of the structural transitions in 2D materials.

**TOC GRAPHIC:**

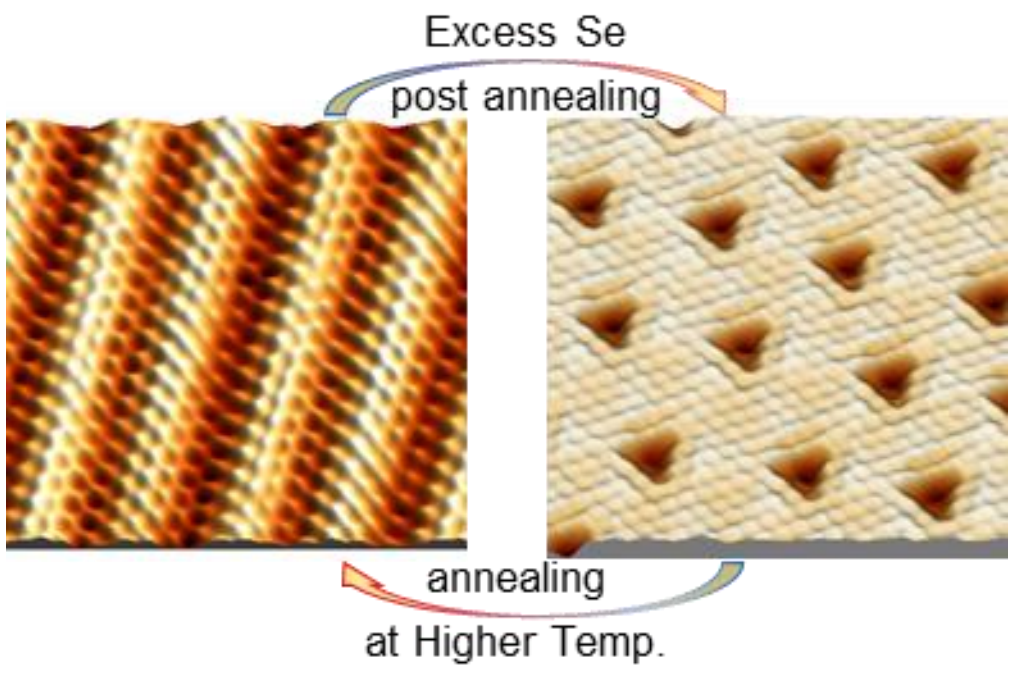

## INTRODUCTION

Structural engineering in two-dimensional (2D) materials has attracted intense interest owing to the significant role in tailoring the structures and properties of nanomaterials.[1,2] Technically, structural engineering in nanomaterials involves the alternations in the atomic arrangement, which could be divided into different catalogues. For instance, tailoring the size or dimensionality is an effective approach to tune the electronic band structures, i.e., cutting graphene into one-dimensional (1D) ribbons giving rise to the formation of a width-dependent band gap.[3,4] Besides, atomic defect engineering like doping[5–7] is also a powerful approach to manipulate the structures and thus the properties of 2D materials. In addition, there could be multiple structural phases with the same atoms or the same chemical formula, such as graphite and diamond. Thus, phase engineering[8,9] provides an alternative way to tune the structures and properties of 2D materials.

Since the first isolation of graphene,[10] metal chalcogenides have provoked interest among physicists and material scientists due to their extraordinary properties and wide applications. For instance, 2D transition metal dichalcogenides (TMDs) are a wide class of 2D materials with sandwiched structures, namely, one layer of transition metal atoms sandwiched by two layers of chalcogen atoms. The properties of TMDs are as diverse as superconductivity,[11–13] quantum spin liquids,[14] and catalytic activity,[15,16] because of the rich combinations of transition metal and chalcogen atoms, as well as the rich structural phases. These diverse novel properties in TMDs thus lead to wide applications in different fields, like optoelectronics,[17,18] sensors,[19–21] energy storage,[22–25] etc. Beyond the TMDs, 2D transition metal monochalcogenides have also attracted much interest owing to their unique structures and properties despite lacking layered bulk counterparts. Very recently, CuSe monolayers have been successfully synthesized through direct selenization of Cu(111) surface.[26–28] Honeycomb structures in CuSe monolayers with different phases were revealed by scanning tunneling microscopy (STM). Direct selenization of Cu(111) at room temperature gives rise to the formation of CuSe monolayers with one-dimensional (1D) moiré patterns. Exotic quantum phenomena like Dirac nodal lines have been unraveled via angular resolved photoelectron spectroscopy (ARPES).[28] In contrast, quasi-ordered arrays of triangular holes are formed in CuSe monolayer by selenization of Cu(111) at elevated temperature, which could be served as templates for nano clusters or self-assembly of organic molecules.[27] Therefore, substrate temperature plays a significant role in the selenization process and could be an effective way to control the structures of CuSe monolayers. However, it is still not clear how the structures change with temperature as well other factors like Se coverage. Thus, systematic study of the selenization of Cu(111) is much need to consistently understand the structural transitions in 2D CuSe.

Here, we report a reversible structural transition in 2D CuSe monolayers on Cu(111) via STM combined with Auger electron spectroscopy (AES). Excess of Se combined with post annealing leads to the structural transition from honeycomb CuSe monolayer with 1D striped pattern to that with quasi-ordered triangular holes, whereas annealing at higher temperature stimulates the reverse transition. Our STM and AES results reveals that both the Se coverage and annealing temperature play essential roles in the formation of 2D CuSe on Cu(111) with different phases, namely the structural transitions.

## EXPERIMENTAL DETAILS

*Sample preparation.* Clean Cu(111) surfaces (single crystal, MaTeck) were prepared by repeated cycles of $Ar^+$ sputtering (600 eV) and annealing at 450 ˚C. The cleanness of the Cu(111) surface was verified by STM topographic images. Se atoms were evaporated through a homemade Knudsen cell at 140 ˚C, while the Cu(111) substrate was kept at room temperature.

All the experiments were carried out in an ultrahigh vacuum (UHV) low temperature (LT) STM system with a base pressure better than $2 \times 10^{-10}$ mbar. The sample was directly transferred to the STM chamber or LEED/AES chamber without breaking the UHV environment after the growth of monolayer CuSe on Cu(111).

*STM measurement.* The STM images were acquired in a constant-current mode at the temperature of liquid nitrogen (~78 K) using an electrochemically etched tungsten tip and all the labelled bias voltages referred to the sample against tip.

*LEED/AES measurement*. The LEED patterns were taken at the electron energy of 90 eV using a LEED/AES system (omicron). The AES spectra were obtained through a LEED/AES system (omicron) with a primary electron- beam energy of 1.5 keV. To minimize the effect of possible inhomogeneity of the samples, all the AES spectra were taken at the same position of samples at different stages.

## RESULTS & DISCUSSION

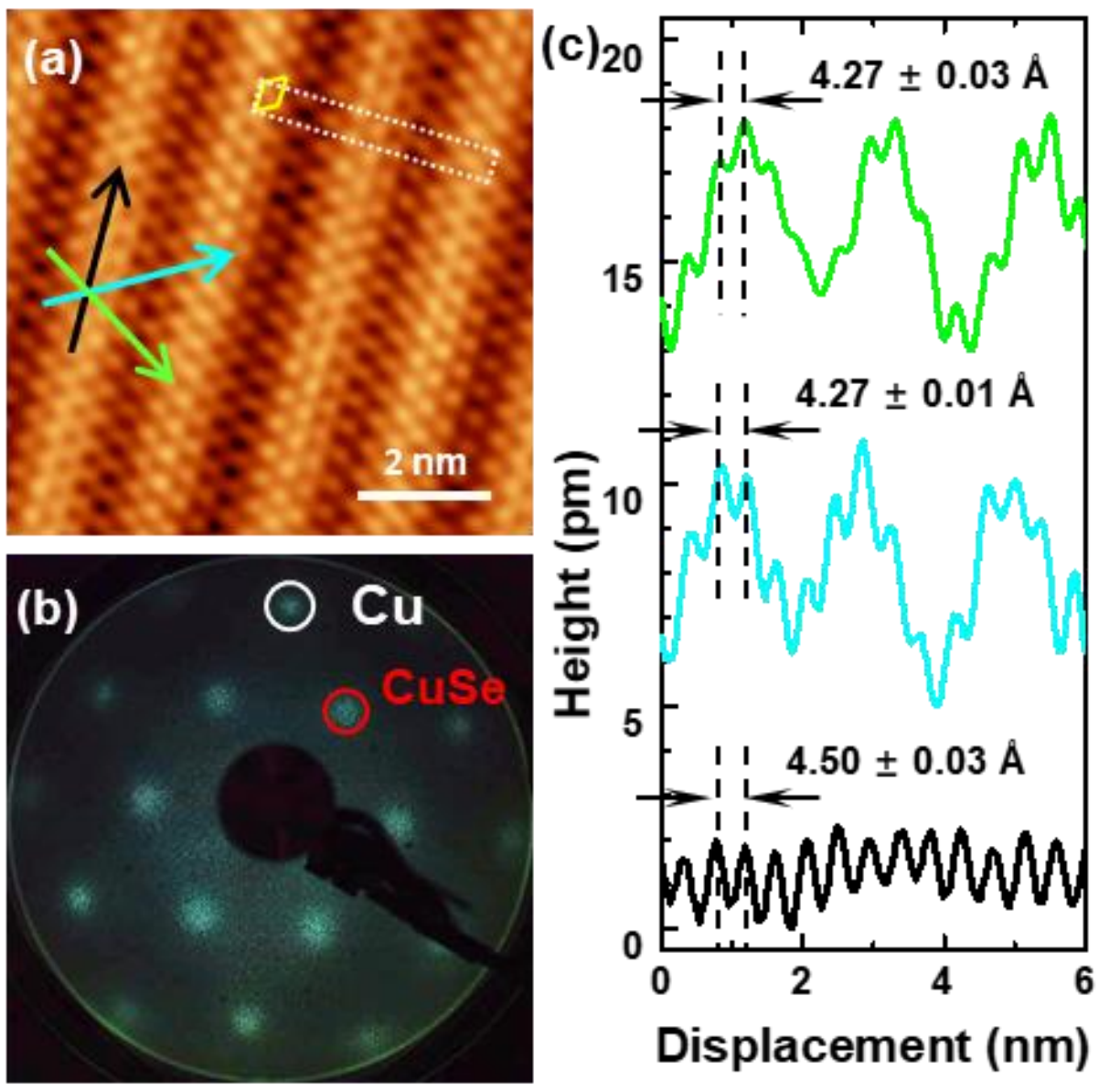

**Figure 1.** Formation of stripe-CuSe on Cu(111) at room temperature. (a) Atomic topographic image and (b) LEED pattern of well-ordered CuSe/Cu(111). (c) Height profiles of CuSe along the high-symmetry directions as indicated by arrows in (a). Tunneling parameters for (a) sample bias U = -0.1 V, tunneling current I = 2.5 nA. LEED pattern in (b) was taken at the energy of 90 eV.

Deposition of Se onto Cu(111) at room temperature gives rise to the formation of well-ordered CuSe monolayers, contrary to the inert Au(111) surface where no chemical reaction with Se even at elevated temperatures.[29] Figure 1a presents a typical STM topographic image of CuSe with atomic resolutions. It reveals the hexagonal lattice of CuSe with average lattice constant of ~4.4 Å, √3-times of that of Cu(111) lattice. Further, the relationship between CuSe and its beneath substrate Cu(111) is confirmed by LEED pattern shown in Fig. 1b, namely, $\sqrt{3}\times\sqrt{3}$-R30° with respect to Cu(111). Besides, it also shows ordered 1D stripes superimposed in the atomic structures, namely, 1D moiré pattern (hereafter referred as stripe-CuSe). These stripes are aligned with the high-symmetry lattice direction, along one of the lattice vectors. Stripes with orientation angle of ±120° respective to the one in Fig. 1a are also found on the same sample. Under certain sample bias, the neighbored two stripes were resolved as one pair (Fig. 2b) or one stripe with period of ~3.6 nm. The unit cell of the 1D moiré pattern is indicated by the white dashed rectangle in Fig. 1a. Height profiles in Fig. 1c reveal the different strains in the high-symmetry directions. The lattice is stretched along the stripe more than the other two directions induced by the lattice mismatch between CuSe and Cu(111) lattices. Along the stripe, the lattice constant was measured to be 4.50 ± 0.03 Å, whereas 4.27 ± 0.03 Å for the other two high-symmetry lattice directions. Thereafter, such lattice distortion results in the formation of 1D moiré pattern. These results are consistent with previous reports.[26,28,30] It should be noted that the 1D moiré structures were not revealed in the LEED pattern in Fig. 1b, which might be due to the relatively low quality of the CuSe thin film and/or Cu substrate in the macroscopic view.

Post-annealing at elevated temperature has been used to manipulate the phase transitions in TMDs. For instance, Lin *et al.*[27] reported that the post-annealing leads to the structural transition from 1T-$PtSe_2$ to patterned 1T/1H-$PtSe_2$ on Pt(111). However, little change in the stripe-CuSe after post-annealing up to 450 ˚C, indicating it is stable even at elevated temperature. Thus, for the stripe-CuSe, post-annealing alone could not lead to the structural transitions and other factors should be taken into account.

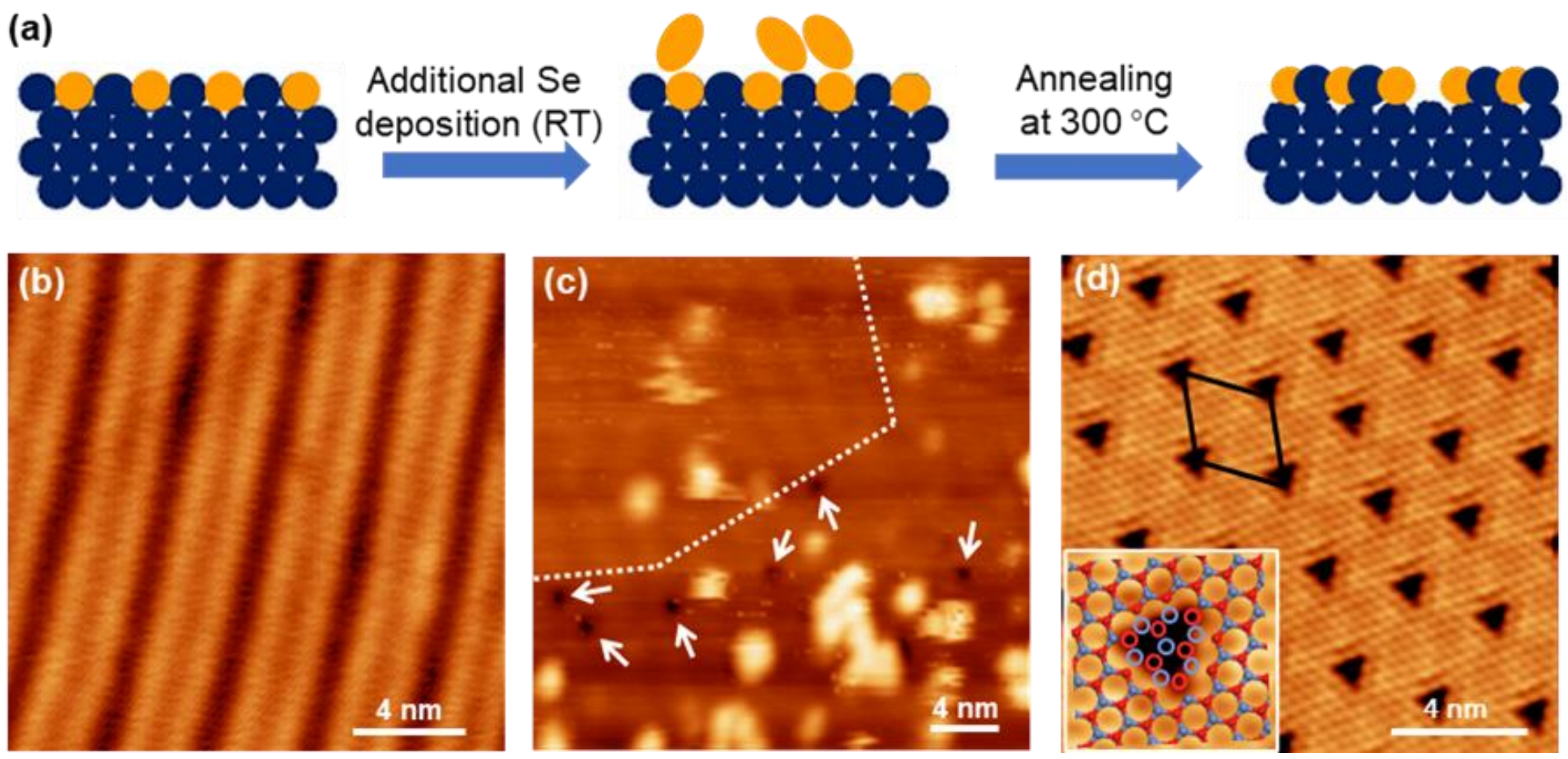

**Figure 2.** Structural transition of CuSe on Cu(111). (a) Schematic process of structural transition from stripe-CuSe to hole-CuSe. (b-d) STM topographic images of CuSe corresponding to each step in the structural transition process in (a). Inset in (d) is the zoomed-in atomic-resolution image overlaid with atomic model for the hole structures, where red and blue balls represent Se and Cu atoms, respectively. The red and blue circles represent the removal Se and Cu atoms in the triangular hole. Tunneling parameters for (a) U = -2.5 V, I = 0.2 nA; (b) U = 3.0 V, I = 0.1 nA; (c) U = 0.2 V, I = 1.0 nA. Samples in (b-d) referred as sample A, B, and C.

In previous works, the Se coverage has significant effects to the structures of CuSe formed on Cu(111) at room temperature.[26] Taking Se coverage into account, we carried out a different post-annealing process to the stripe-CuSe. As shown in Fig. 2a, prior to the post annealing, additional Se were deposited onto the sample surface at room temperature. Surprisingly, Figures 2b and 2c reveal that further deposition of Se onto the stripe-CuSe/Cu(111) results in dramatic change in the topography. Besides of the stripe-CuSe regions, it shows some regions with flat surface along with small holes indicated by the white arrows in Fig. 2c. In addition, more clusters of Se atoms or Se-Cu alloys were found on the flat-surface regions than stripe-CuSe regions, which might imply the stripe-CuSe surface with higher mobility for the atoms or clusters.

Interestingly, further annealing of this sample at 150~300 ˚C leads to the formation of a different phase of CuSe with quasi-ordered arrays of triangular holes with overall period of ~3 nm (Fig. 2d, hereafter referred as hole-CuSe), which has been reported in previous work by direct selenization of Cu(111) at elevated temperature.[27,31] The triangular holes could be modelled as the removal of 3 honeycomb rings, namely, the removal of 13 atoms including 6 Se atoms and 7 Cu atoms, as shown in the inset of Fig. 2d. Besides, it reveals little distortions in the CuSe lattices with lattice constant of 4.13 ± 0.03 Å, implying the lattice mismatch induced strains are released through the formation of triangular holes in the lattice.

So far, we have demonstrated the structural transition from stripe-CuSe to hole-CuSe involving Se coverage as well as post annealing. It is naturally to ask whether such transition is reversible, namely, whether the structural transition from hole-CuSe to stripe-CuSe could occur under certain conditions. Interestingly, we found that post annealing at higher temperature stimulates the reverse transition, namely, from hole-CuSe to stripe-CuSe.

Figures 3a and 3b present the STM topographic images after further annealing hole-CuSe/Cu(111) at 400 ˚C and 450 ˚C, respectively. After annealing at 400 ˚C, only a few of the triangular holes were found in CuSe/Cu(111), as shown in Fig. 3a. Instead, it shows quasi-ordered networks in the vicinity of these holes. We note that the hole-CuSe has been reported to be ambient stable and stable in vacuum up to high temperature (400 ˚C) in previous work,[27] probably because the exact substrate temperature was lower than that in the present work. Further, Figure 3b reveals such structures is converted to well-ordered stripe structures after further annealing the sample at 450 ˚C. The periodicity of such stripe structure was measured to be ~2.9 nm, which is around 20% smaller that in stripe-CuSe formed by direct selenization of Cu(111) at room temperature. Each of these stripes is resolved as a pair of stripes in the atomical-resolution STM image (Fig. 3c), the same as those with larger period of ~3.6 nm. Besides, these tripes are parallel to one of the high-symmetry lattice directions of CuSe, which is also confirmed by the parallel reciprocal lattice vectors of the stripe structures and atomic lattice of Cu(111) in the fast Fourier transform image in Fig. 3d. Along the stripe, the lattice constant of CuSe was measured to be 4.50 ± 0.03 Å, whereas 4.25 ± 0.02 Å and 4.23 ± 0.03 Å for the other two high-symmetry lattice directions with slight difference compared with those for stripe-CuSe with period of ~3.6 nm.

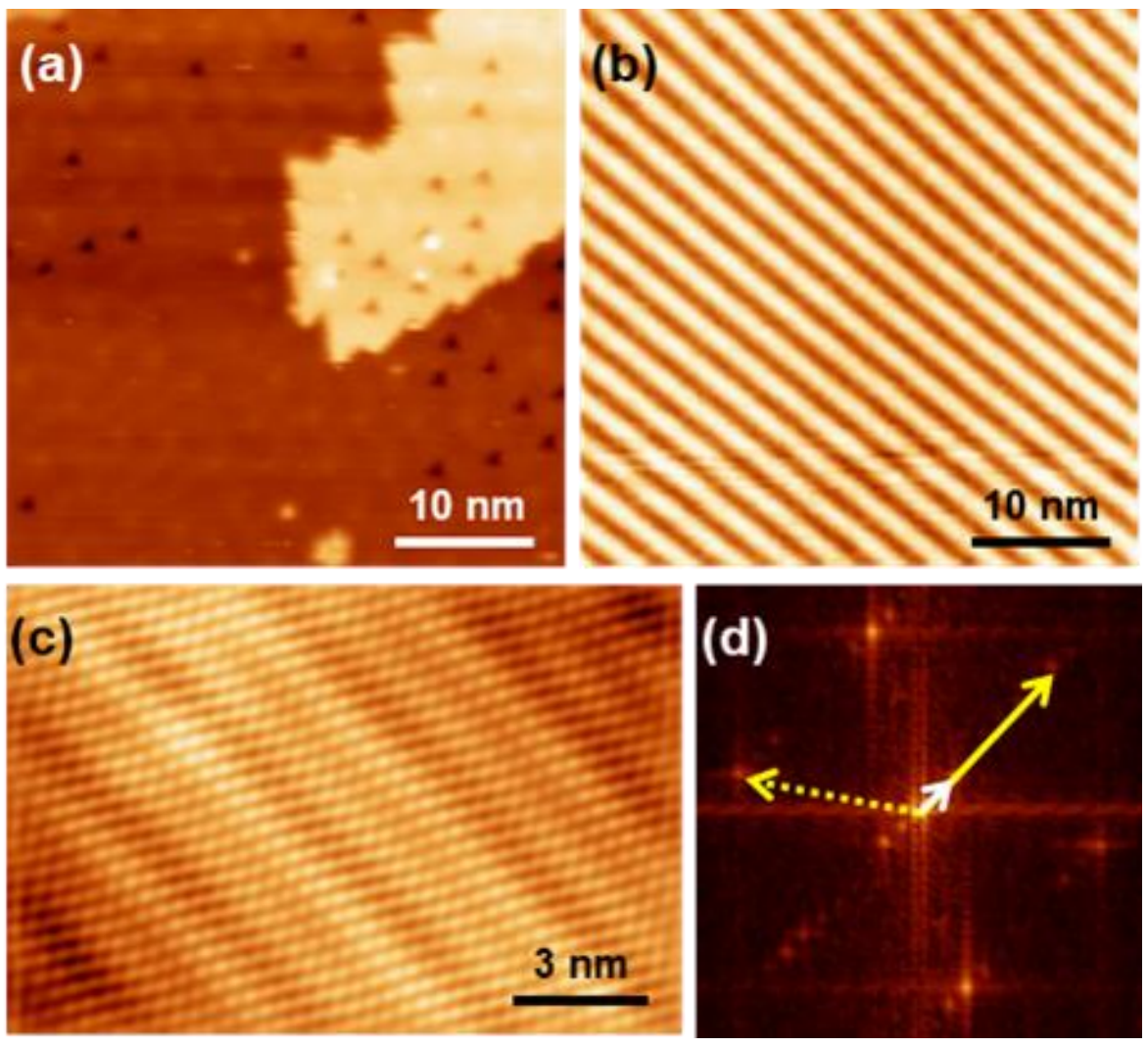

**Figure 3.** Reverse structural transition of CuSe on Cu(111). (a) and (b) STM topographic images after post annealing of hole-CuSe at (a) 400 ˚C and (b) 450 ˚C, respectively. (c) Atomic-resolution STM topographic and (d) corresponding FFT images of the stripe structure of CuSe in (b). The white arrow in (d) denotes the reciprocal lattice vector for 1D stripe structure, whereas the yellow arrows for the hexagonal CuSe, where the solid one is parallel to the white arrow. Tunneling parameters for (a) U = -2.5 V, I = 0.1 nA; (b) U = 1.0 V, I = 0.1 nA; (c) U = 5 mV, I = 2.5 nA. Sample in (b) is referred as sample D.

To verify such slight difference in the lattice distortions could lead to the formation of 1D stripe structures with different periodicities, we built structure models for CuSe/Cu(111) based on the STM results, as shown in Fig. 4. Considering the lattice distortions in CuSe, the unit cell of CuSe is a rectangle consisting of two Se and two Cu atoms rather than a rhombus consisting of one Se and one Cu atoms, as shown in Fig. 4a. The lattice constants for these distorted lattices of CuSe are 7.26 Å, 4.5 Å, and 7.16 Å, 4.5 Å, respectively. Simply stacking of such CuSe lattices with Cu(111) lattice well reproduces the 1D stripe structures with periodicities of 3.6 nm and 2.9 nm, respectively, in good agreement with the STM results. The unit cells of the 1D stripe structures are 5×1-CuSe on 12×√3-Cu(111) and 4×1-CuSe on 11×√3-Cu(111), respectively. Therefore, the stripe-CuSe with different periods originate from the lattice mismatch induced strain in CuSe, which results in asymmetric distortions in the CuSe lattices. It is worth to note that the stripe-CuSe with period of 2.9 nm could also be converted to the hole-CuSe through the same process as illustrated in Fig. 2a. Such structural transition implies overall the same atomic structures in the stripe-CuSe despite of different periodicities.

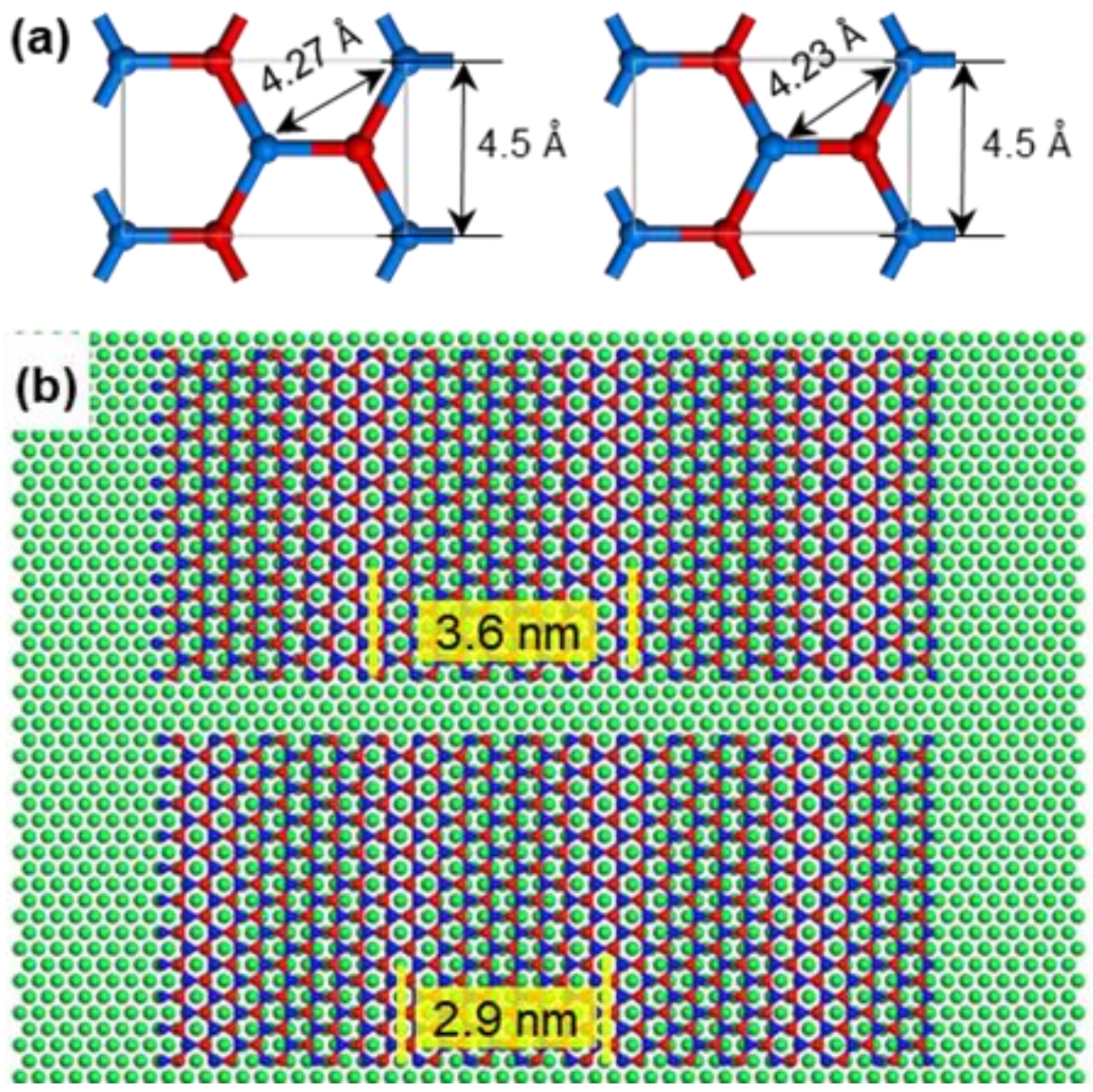

**Figure 4.** Structural models of CuSe on Cu(111) with different lattice distortions. (a) Unit cells of CuSe with lattice distortions. (b) CuSe on Cu(111) with 1D stripe structures by simply

stacking of CuSe sheets and Cu(111) sheet (green balls). The lattice distortions are based on the STM results, and the structures are not optimized.

So far, we have illustrated the reversible structural transition in CuSe monolayers on Cu(111): the stripe-CuSe and hole-CuSe. With additional Se plus post annealing, the stripe-CuSe could be converted to the hole-CuSe, whereas post annealing of the hole-CuSe at higher temperature gives rise to the formation of stripe-CuSe on Cu(111). Therefore, both the Se coverage and post annealing play important roles in such structural transitions. With this aspect, we carried out AES measurement for the CuSe samples with different phases to identify the relative content of Se in these samples.

Figure 5a presents the AES spectra for CuSe samples with different phases as well as clean Cu(111) for reference. For clean Cu(111), the spectrum shows a pair of peak and dip at 102.8 eV and 109.2 eV, respectively. For CuSe samples, the peak shifts to lower energy (~102.1 eV) and the dip remains almost the same, which could be explained by the formation of chemical bonds Cu-Se on the surface. Besides, a pair of peak and dip at 92.5 eV and 98.7 eV are shown in the spectra, which are associated to the Se on the surface. The relative AES intensities of Se for CuSe samples are plotted in Fig. 5b, which could qualitatively represent the relative Se concentration on the surface. It clearly shows the increase in Se AES intensity after additional Se deposition: from 0.29 for sample A to 0.51 for sample B. Further annealing leads to a slight increase in the Se AES intensity in sample C (hole-CuSe), implying that the slight increase of the Se/Cu ratio on the surface. Such slight increase of Se/Cu ratio is consistent with the formation of triangular holes, where more Cu atoms are removed than Se atoms in CuSe lattices. The lower Se AES intensity in sample D compared with sample C suggests that annealing at higher temperature gives rise to the desorption of Se or diffusion of Se atoms into the bulk of Cu substrate. Therefore, both the Se coverage and annealing temperature play significate roles in the reversible structural transition in CuSe on Cu(111).

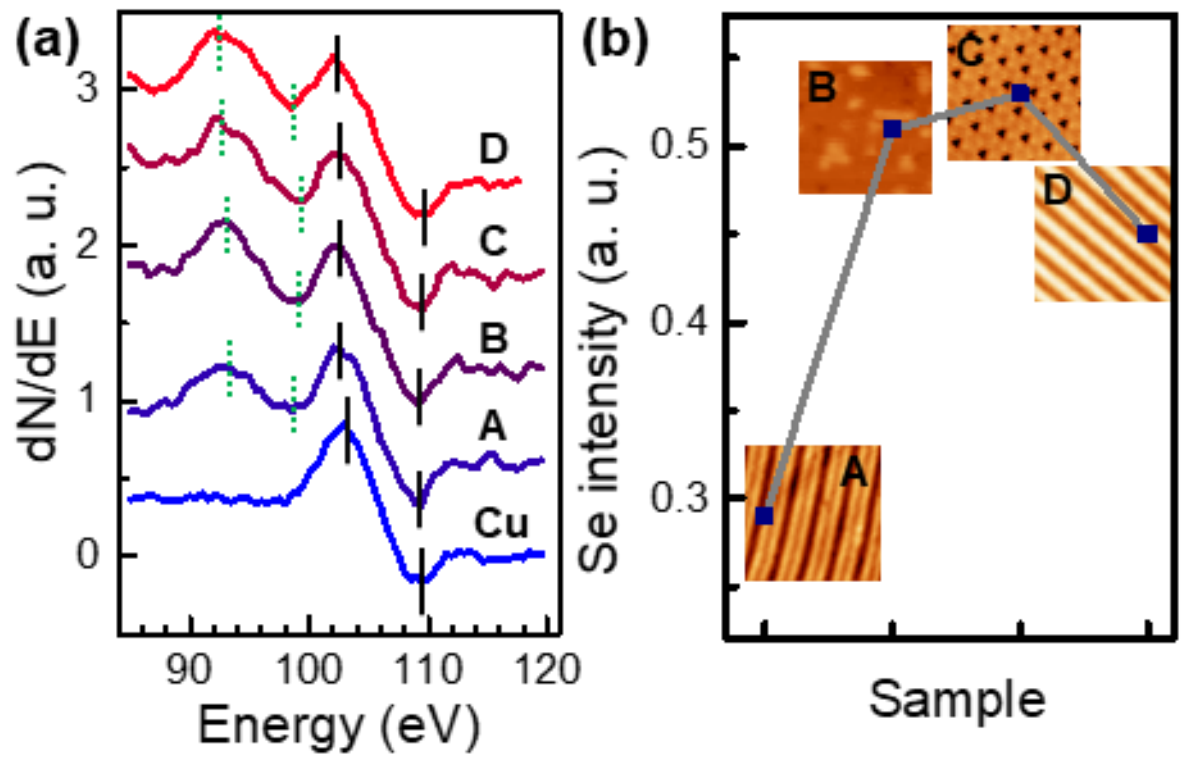

**Figure 5.** Auger electron spectroscopy of CuSe/Cu(111). (a) AES spectra of CuSe with different structures. The AES peaks for Cu and Se elements are denoted by solid and dashed lines, respectively. All the spectra are normalized by the peak-peak intensity of Cu at ~110 eV. The label A-D referred to samples A-D which are defined in Figs. 2 and 3. (b) AES intensity of Se for these samples.

It should be noted that the Se concentration in the hole-CuSe structures is slightly smaller than that of the stripe-CuSe structures according to the structural models, which is controversial the AES results as well as the fact that transition from the stripe-CuSe to hole-CuSe requires excess Se deposition. To explain such controversy between the theoretic models and experimental results, we speculate that the extra Se atoms serve as interfacial or sub-surface Se in the hole-CuSe/Cu(111) system. The interfacial or sub-surface Se might affect the formation energy of CuSe as well as the dynamic process during the growth, and thus change the final structures of CuSe. Similar phenomena have been observed in other systems. For instance, the existence of interfacial oxygen determines the edge orientations of MgO islands on Ag(100).[32] Under Se-poor conditions, the system favors favor the formation of CuSe with 1D moiré structures, namely, stripe-CuSe. In contrast, under Se-rich conditions, extra Se atoms serve as interfacial or sub-surface Se, which favors the formation of CuSe with triangular hole structures, namely, hole-CuSe. The Se content could be reversibly manipulated between Se-poor and Se-rich conditions by excess Se deposition plus post annealing and/or annealing at high temperatures, leading to the reversible structural transition of CuSe. Besides, the formation of 1D moiré structure with different periodicity after post-annealing of hole-CuSe at high temperature could also be explained by the existence of interfacial or sub-surface Se, which could be supported by the higher Se intensity in AES results (Fig. 5).

## CONCLUSIONS

In summary, we have demonstrated a reversible structural transition in 2D CuSe on Cu(111) through in-situ STM and AES. Direct selenization of Cu(111) gives rise to the formation of stripe-CuSe, namely, honeycomb CuSe monolayer layers with 1D moiré structures, due to the asymmetric lattice distortions in CuSe induced by the lattice mismatch. Additional deposition of Se combined with post annealing results in the structural transition from the stripe-CuSe to hole-CuSe with quasi-ordered arrays of triangular holes. Further, annealing the hole-CuSe at higher temperature leads to the reverse structural transition, namely from hole-CuSe to stripe-CuSe. AES measurement unravels the Se content change in such reversible structural transition. Therefore, both the Se coverage and annealing temperature play significant roles in the reversible structural transition in CuSe on Cu(111). Our work provides insights in understanding of the structural transitions in 2D materials.

## AUTHOR INFORMATION

**Corresponding Authors**

*Email: ydque@phy.cuhk.edu.hk (Y.D.Q.).

*Email: xdxiao@phy.cuhk.edu.hk (X.D.X.).

**ORCID**

Yande Que: 0000-0002-5267-4985

Chaoqiang Xu: 0000-0001-8561-8974

Yuan Zhuang: 0000-0003-3634-8546

Xudong Xiao: 0000-0003-0551-1144

**Author Contributions**

Y.Z. and Y.D.Q. Contributed equally to this work.

**Notes**

The authors declare no competing financial interest.

## ACKNOWLEDGEMENTS

This work was supported by the Research Grant Council of Hong Kong (No. 404613), the Direct Grant for Research of CUHK (No. 4053306 and No. 4053348).